# A NEW TREND IN OPTIMIZATION ON MULTI OVERCOMPLETE DICTIONARY TOWARD INPAINTING

*SeyyedMajid Valiollahzadeh*[(a)], *Mohammad Nazari*[(b)], *Massoud Babaie-Zadeh*[(a)], *Christian Jutten*[(c)]

[(a)]Department of Electrical Engineering, Sharif University of Technology, Tehran, Iran
[(b)]Department of Electrical Engineering, Amirkabir University of Technology, Tehran, Iran
[(c)]Laboratoire des Images et des Signuax, Institute National Polytechnique de Grenoble, France

**ABSTRACT**

Recently, great attention was intended toward overcomplete dictionaries and the sparse representations they can provide. In a wide variety of signal processing problems, sparsity serves a crucial property leading to high performance. Inpainting, the process of reconstructing lost or deteriorated parts of images or videos, is an interesting application which can be handled by suitably decomposition of an image through combination of overcomplete dictionaries. This paper addresses a novel technique of such a decomposition and investigate that through inpainting of images. Simulations are presented to demonstrate the validation of our approach.

## 1. INTRODUCTION

Sparse signal decomposition of signals on an over overcomplete Dictionary was of great interest among researchers in past few years and serve many interesting applications [1]. The main assumption over these signals is that they are linear mixtures of building atoms and also only a few of these atoms will participate in the reconstruction. In the context of image processing an interesting decomposition application would be separating texture from non-texture part to be used in areas from compression to analysis and synthesis of an image[2][3].

Inpainting consists in problems like filling the holes, reconstructing lost or deteriorated parts of images or videos, removal of scratches in old photos, removal of unwanted text or graphic and is an interesting inverse problem with lots of research momentum [4] in recent years dealing highly with such decomposition. Pioneered by the work of Sapiro et al [5], total variation was used in this respect taking mainly the geometrical contents into consideration. Since images contain both geometrical and textural information, decomposition should be done in two layers.

This approach has been presented in [6]-[7] and the layers to which an image is decomposed are called texture and cartoon. The inpainting process is done in each layer separately and afterwards the output will be formed by summing up these layers. The crucial part in this approach is layer decomposition and will extend the notions of total variations. By this trend, if any failure in the inpainting of each layer is presented, superimposing of two layers will lead in less visual artifact and hence quite satisfactory result.

In some recent work sparsity was taken into account as additional criteria to decompose an image to these layers. To this end, we need two dictionaries, mutually incoherent, one to represent the texture and the other for the cartoon. Both should provide the sparse representation for the corresponding layer image while yielding nonsparse for the other. Combination of these two dictionaries into one and performing the (Basis Pursuit denoising) BPDN [1] algorithm seeking the sparsest solution has shown to perform well and even can be improved by further applying the total-variation regularization.

Elad *et al*. [8, 9] proposed an inpainting algorithm capable of filling in holes in either texture or cartoon content, or any combination thereof extending employment of separation by sparsity, so that the missing samples fit naturally into the layer separation framework. The main advantageous point of this approach is the global treatment trend toward the image rather the local one. Also it deploys general overcomplete dictionaries which can be better established for a typical image content.

What is presented in this paper is quite similar on the basis of sparse representations, but modeling the overall problem as a specific optimization is better relaxed. Inspired by the work of Mohimani, *et al*. [10] for finding

---

[1] This work has been partially supported by Iran National Science Foundation (INSF) under contract number 86/994, by Iran Telecommunications Research Center (ITRC), and also by ISMO and French embassy in Iran in the framework of a Gundi-Shapour collaboration program

the sparsest solution of an Underdetermined System of Linear Equations (USLE) through the smoothed $\ell^0$-norm, we extend this approach in two dimensional models to solve the prior modeling. The outline of the paper is as follows. In section 2, we briefly present the modeling scenario to decompose a signal over two incoherent dictionaries. In section 3 we model the inpainting problem and present the final algorithm. We discuss some simulation results to validate the proposed algorithm in section 4 and finally conclusion and summary of later work is discussed in the last section.

## 2. MAIN IDEA

Let the input image $\mathbf{c}$ containing $N$ total pixels, be presented as a one-dimensional vector. This image is to be decomposed over two distinct dictionaries, $\mathbf{A}$ and $\mathbf{B}$, the former corresponding to texture and the latter to cartoon. Both provide sparse representation for the image of their kind and non-sparse for the other, written formally as:

$$\mathbf{c}_1 = \mathbf{A}\mathbf{s}_1 \quad (\mathbf{s}_1 \text{ is sparse}) \quad (1)$$
$$\mathbf{c}_2 = \mathbf{B}\mathbf{s}_2 \quad (\mathbf{s}_2 \text{ is sparse}) \quad (2)$$

Sparsity of a vector S is quantified by its $\ell^0$-norm, denoted by $\|\mathbf{s}\|_0$, defined by the number of its nonzero elements. There are two assumptions over these dictionaries [8,9]: firstly, these two dictionaries should be incoherent, i.e. the texture dictionary is not able to represent the cartoon image sparsely and vice versa. Secondly, the dictionary assigned to texture should be such that if the texture appears in parts of the image and is otherwise zero, representation is still sparse, implying somehow that it should employ a multiscale and local analysis of the image content.

Now, we seek a sparse representation over the combined dictionary:

$$\{\mathbf{s}_1, \mathbf{s}_2\} = \underset{\mathbf{s}_1, \mathbf{s}_2}{\operatorname{argmin}} \left\{ \|\mathbf{s}_1\|_0 + \|\mathbf{s}_2\|_0 \right\} \quad (3)$$

$$\text{Subject to:} \quad \mathbf{A}\mathbf{s}_1 + \mathbf{B}\mathbf{s}_2 = \mathbf{c}$$

The problem is non-convex and seemingly intractable due to combinatorial search it needs, however inspired by the work of Mohimani et al [10], we can find $\mathbf{s}_1, \mathbf{s}_2$ as it using smoothed $\ell^0$-norm. Smoothed $\ell^0$-norm of a vector $\alpha$ is an approximation to its $\ell^0$-norm and is defined as:

$$F_\sigma(\boldsymbol{\alpha}) = \sum_{i=1}^{m} \exp(-\alpha_i^2 / 2\sigma^2) \quad (4)$$

where $\alpha$ is a parameter determining a tradeoff between the accuracy of approximation and the smoothness of $F_\sigma(\boldsymbol{\alpha})$. Minimizing the $\ell^0$ norm of $\boldsymbol{\alpha}$ subject to $\mathbf{b} = \boldsymbol{\Phi}\boldsymbol{\alpha}$ then requires then to maximize $F_\sigma(\boldsymbol{\alpha})$ for a small value of $\sigma$. For a small $\sigma$, $F_\sigma(\boldsymbol{\alpha})$ is highly non-smooth with lots of local maxima. To overcome this difficulty we use a decreasing sequence of $\sigma$ and make use of maximizer of $F_\sigma(\boldsymbol{\alpha})$ as a starting point to find the next (smaller) sigma [10]. Moreover, the algorithm initially starts with minimum $\ell^2$ norm solution of $\mathbf{b} = \boldsymbol{\Phi}\boldsymbol{\alpha}$, which corresponds to the maximizer of $F_\sigma(\boldsymbol{\alpha})$ when $\sigma \to \infty$.

Using similar idea, we want to minimize a cost function $J_\sigma(\mathbf{s})$ -which will be introduced in the next section- subject to $\mathbf{A}\mathbf{s}_1 + \mathbf{B}\mathbf{s}_2 = \mathbf{c}$. The minimization should be done for small $\sigma$ and in order to avoid trapping in local minima we use a sequence of $\left[\sigma_1, ..., \sigma_{k_{\max}}\right]$ and then minimize $J_\sigma(\mathbf{s})$ for each $\sigma$, with the starting point yielded by the maximizer of the previous (longer) $\sigma$. Moreover the process is initialized by:

$$\begin{bmatrix} \mathbf{A} & \mathbf{B} \end{bmatrix} \begin{bmatrix} \mathbf{s}_1 \\ \mathbf{s}_2 \end{bmatrix} = \mathbf{c} \Rightarrow$$
$$\begin{bmatrix} \mathbf{s}_1 \\ \mathbf{s}_2 \end{bmatrix} = \begin{bmatrix} \mathbf{A} & \mathbf{B} \end{bmatrix}^\dagger \mathbf{c} = \begin{bmatrix} \left(\mathbf{P}_\mathbf{B}^\perp \mathbf{A}\right)^\dagger \\ \left(\mathbf{P}_\mathbf{A}^\perp \mathbf{B}\right)^\dagger \end{bmatrix} \mathbf{c} \quad (5)$$

where, $\mathbf{P}_\mathbf{A}^\perp$ and $\mathbf{P}_\mathbf{B}^\perp$ are the orthogonal projections of the corresponding matrices:

$$\mathbf{P}_\mathbf{A}^\perp = \mathbf{I} - \mathbf{A}^T \left(\mathbf{A}\mathbf{A}^T\right)^{-1} \mathbf{A}$$
$$\mathbf{P}_\mathbf{B}^\perp = \mathbf{I} - \mathbf{B}^T \left(\mathbf{B}\mathbf{B}^T\right)^{-1} \mathbf{B} \quad (6)$$

Then we use L iterations of the steepest ascent algorithm, followed by a projection onto the feasible which is:

$$\text{Update} \begin{bmatrix} \mathbf{s}_1 \\ \mathbf{s}_2 \end{bmatrix} = \begin{bmatrix} \mathbf{s}_1 \\ \mathbf{s}_2 \end{bmatrix} - \begin{bmatrix} \mathbf{A} & \mathbf{B} \end{bmatrix}^\dagger (\mathbf{c} - \mathbf{A}\mathbf{s}_1 - \mathbf{B}\mathbf{s}_2) \quad (8)$$

For more explanation about choosing the sequence see [10].

## 3. Modeling inpainting and the final proposed algorithm

Suppose that missing pixels of the image are masked with a diagonal mask matrix $\mathbf{M}$ (of which has value '1' over the existing pixels and '0' over the missing pixels) we propose restoring the image by optimizing the following problem:

$$\{\mathbf{s}_1^{Opt}, \mathbf{s}_2^{Opt}\} = \arg\min_{\mathbf{s}_1, \mathbf{s}_2} \left\{ \|\mathbf{s}_1\|_0 + \|\mathbf{s}_2\|_0 + \lambda \|\mathbf{M}(\mathbf{c} - \mathbf{A}\mathbf{s}_1 - \mathbf{B}\mathbf{s}_2)\|_2^2 + \gamma \text{TV}\{\mathbf{A}\mathbf{s}_1\} \right\} \quad (9)$$

in which we have $TV\{\mathbf{x}\} = \|\nabla \mathbf{x}\|_1$. So the recovered image would be:

$$\hat{\mathbf{c}} = \mathbf{A}\mathbf{s}_1^{Opt} + \mathbf{B}\mathbf{s}_2^{Opt} \qquad (10)$$

---

− *Initialization* :

+ *Let* $\begin{bmatrix} \mathbf{s}_1 \\ \mathbf{s}_2 \end{bmatrix} = \begin{bmatrix} \mathbf{A} & \mathbf{B} \end{bmatrix}^\dagger \mathbf{c}$

+ *Choose a suitable sequence for* $\sigma = \begin{bmatrix} \sigma_1, ..., \sigma_{N_m} \end{bmatrix}$

− *For* $n = 1, ..., N$

   + *Maximize the function* $J_\sigma(\mathbf{S})$ *using steepest descent algorithm*

   − *For* $k = 1, ..., L$

- $\begin{bmatrix} \Delta \mathbf{s}_1 \\ \Delta \mathbf{s}_2 \end{bmatrix} =$

$\begin{bmatrix} s_{11}^{(n)} \exp\left(-(s_{11})^2 / 2\sigma_n^2\right) & \cdots & s_{1m_1}^{(n)} \exp\left(-(s_{1m_1})^2 / 2\sigma_n^2\right) \\ s_{21}^{(n)} \exp\left(-(s_{21})^2 / 2\sigma_n^2\right) & \cdots & s_{2m_2}^{(n)} \exp\left(-(s_{2m_2})^2 / 2\sigma_n^2\right) \end{bmatrix}$

$+ 2\lambda \begin{bmatrix} \mathbf{A}^T(\mathbf{c} - \mathbf{A}\mathbf{s}_1) \\ \mathbf{B}^T(\mathbf{c} - \mathbf{B}\mathbf{s}_2) \end{bmatrix}.$

- $\begin{bmatrix} \mathbf{s}_1 \\ \mathbf{s}_2 \end{bmatrix} = \begin{bmatrix} \mathbf{s}_1 \\ \mathbf{s}_2 \end{bmatrix} - \begin{bmatrix} \mu_1 \Delta \mathbf{s}_1 \\ \mu_2 \Delta \mathbf{s}_2 \end{bmatrix}$

($\mu_1$ and $\mu_2$ can be chosen by a line-search minimizing the overal penalty function or fixed stepsize.)

- *Calculate* $\mathbf{c}_1 = \mathbf{A}\mathbf{s}_1$

- *Apply the TV Corrolation on the* $\mathbf{c}_1$

   − *Reconstruct* $\mathbf{c}_1 = \mathbf{c}_1 - \mu \dfrac{\partial TV\{\mathbf{c}_1\}}{\partial \mathbf{c}_1}$

             $= \mathbf{c}_1 - \mu \nabla \left(\dfrac{\nabla \mathbf{c}_1}{|\nabla \mathbf{c}_1|}\right)$

- *Update* $\lambda = \lambda - \dfrac{1}{N}\lambda_{\max}$

- *Update* $\begin{bmatrix} \mathbf{s}_1 \\ \mathbf{s}_2 \end{bmatrix} = \begin{bmatrix} \mathbf{s}_1 \\ \mathbf{s}_2 \end{bmatrix} - \begin{bmatrix} \mathbf{A} & \mathbf{B} \end{bmatrix}^\dagger (\mathbf{c} - \mathbf{A}\mathbf{s}_1 - \mathbf{B}\mathbf{s}_2)$

+ $s = s_{n-1}$

− *Final coefficients are* $\mathbf{s}_1$ *and* $\mathbf{s}_2$.

Figure 1: The proposed algorithm for decomposition

---

The term $TV\{\mathbf{A}\mathbf{s}_1\}$ essentially computes the image $\mathbf{A}\mathbf{s}_1$ (supposed to be piecewise smooth), applies the absolute gradient field and summing up $\ell^1$ norm to avoid blockiness and force the image be smooth thus support the separation process.

These coefficients to be found can be relaxed as stated in the previous part:

$$J_\sigma = \left(M_1 - F_\sigma(\mathbf{s}_1)\right) + \left(M_2 - F_\sigma(\mathbf{s}_2)\right) \\ + \lambda \|\mathbf{M}(\mathbf{c} - \mathbf{A}\mathbf{s}_1 - \mathbf{B}\mathbf{s}_2)\|_2^2 + \gamma TV\{\mathbf{A}\mathbf{s}_1\} \qquad (11)$$

$M_1$ and $M_2$ are the length of $\mathbf{s}_1, \mathbf{s}_2$ coefficients, not necessarily equivalent. The overall algorithm is shown in Fig 1. The parameters $\gamma$ and $\lambda$ are found experimentally [9].

## 4. EXPERIMENTAL RESULTS

In this section, we apply the algorithm of Fig1 for the reconstruction of gray level still images where some parts are missing. In proposed algorithm, we briefly present the scenario to decompose a signal over two incoherent dictionaries. Our approach in this work is to choose two known transforms, one to represent the texture and the other for the cartoon.

With regards to the actual choice, for the cartoon representation, we used curvelet transform and for the texture; we used local-DCT transform. These dictionaries are nice choice of transform according to our experience dependent on this problem. We must remind that type of sparse transformation may vary from one image to another [8] but must be mutually independent.

In fig 2, we show the representation result of the proposed algorithm for the Barbara image. Left image was obtained using the curvelet transform with six resolution levels and right one is the output of local-DCT representation with a block size 32×32. We must mention that resolution levels in curvelet and optimal block size in local-DCT transformation were obtained experimentally.

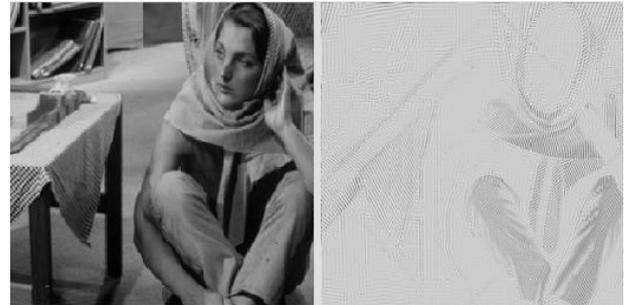

Figure 2: The representation result in last iteration of proposed algorithm for the Barbara image.(left) Output of curvelet transform with six resolution levels. (right) Output of local-DCT representation with a block size 32×32.

The parameters we had used in our simulations are: $N=5$ (number of decreasing value of $\sigma$), $\lambda_{max} \in [1,2]$ and $L=10$ (number of iterations of the steepest ascent algorithm). Note that for calculating the computational complexity of the proposed inpainting algorithm, we can ignore L iterations of the steepest ascent calculation, therefore it is governed by the number of applying the two forward and the inverse transforms.

In fig 2, (top left) we show the original Barbara image; on top right the target regions are masked in white. Region filling via our inpainting method using curvelet and local-DCT dictionaries are illustrated on bottom left. The result of our algorithm around Barbara's eyes shows no trace of the original holes, and seems natural on bottom right.

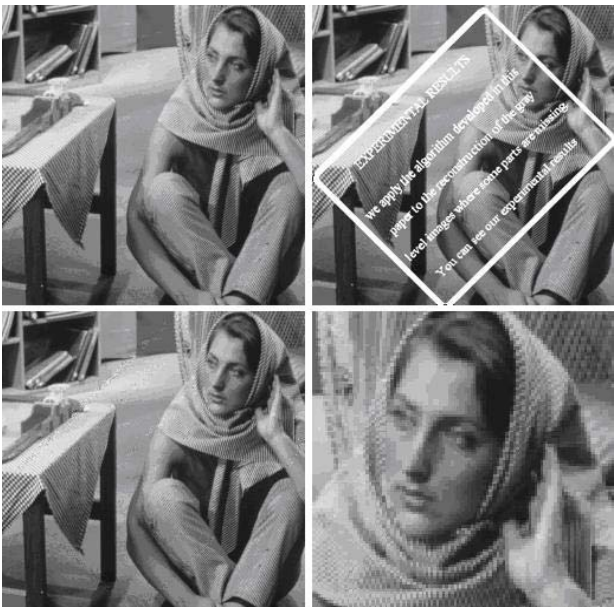

Figure 2: The reconstruction of the masked image.
(top left) Original image. (top right) The target regions are masked in white. (bottom left) Region filling via the proposed inpainting algorithm. (bottom right) The result of our algorithm around Barbara's eyes.

## 5. CONCLUSIONS

In this paper we presented a novel approach for inpainting. It is basically on the basis of decomposition of an image to texture and cartoon layers via sparse combinations of atoms of predetermined dictionaries. The stated algorithm with consideration of total-variation regularization attempts to fill in the holes in each layer separately and superimposes these layers as a final solution. Experimental results show the efficiency of the proposed algorithm in finding the missing samples. Future theoretical work on the general behaviour of this algorithm along with learning of dictionaries through examples adapted to each layers are two further topics in our current research agenda.